\documentclass[11pt]{article}

\usepackage[T1]{fontenc}
\usepackage[utf8]{inputenc} 
\usepackage{lmodern}
\usepackage[margin=1in]{geometry}
\usepackage{microtype}

\usepackage{amsmath, amssymb, amsfonts}
\usepackage{bm}
\usepackage{siunitx}

\usepackage{graphicx}
\usepackage{booktabs}
\usepackage{multirow}
\usepackage{array}
\usepackage{caption}
\usepackage{subcaption}
\usepackage{float}

\usepackage[super,sort&compress]{natbib}

\usepackage[colorlinks=true,linkcolor=blue,citecolor=blue,urlcolor=blue]{hyperref}

\title{\textbf{Measuring Elastic Properties of Granular Hydrogels: Effects of Capillary Interaction and Ionic Conditions}}

\author{
Jiayin Zhao$^{a}$, Haiyi Zhong$^{a}$, Yixiang Gan$^{a,*}$\thanks{Corresponding author: yixiang.gan@sydney.edu.au}
\\[4pt]
\small $^{a}$School of Civil Engineering, The University of Sydney, Sydney, NSW 2006, Australia
}
\date{}

\begin{document}
\maketitle

\begin{quote}
\noindent
Abstract: The elastic properties of granular hydrogels are commonly characterised under wet conditions, yet the influence of capillary interactions remains unclear. In practical applications, hydrogels operate in aqueous environments containing dissolved ionic species, where swelling and elastic behaviour depend sensitively to ionic conditions. In this study, an experimental setup is developed to measure elastic responses of granular hydrogels under wet conditions and to directly observe liquid bridge formation and evolution during compression. Results show that neglecting capillary contributions leads to a systematic underestimation of the Young's modulus, with a size-dependent deviation. In addition, the measured elastic response exhibits ionic-condition-dependent behaviour consistent with swelling-dependent constitutive trends. These findings highlight the necessity of separating capillary and mechanical contributions when characterising soft granular materials in wet environments.
\end{quote}

\section{Introduction}
Hydrogels are cross-linked polymer networks capable of substantially absorbing aqueous substances leading to their tuneable properties, and have been broadly and rapidly adapted in various engineering applications, such as environmental engineering \cite{ref1,ref2}, soft robotics \cite{ref3,ref4} and biomedical engineering \cite{ref5,ref6}. For example, owing to their high hydration levels and tuneable elastic properties, hydrogels demonstrate good biocompatibility and are therefore widely used in tissue engineering, drug delivery, and wound care \cite{ref7,ref8,ref9}. Given that hydrogels interact with living tissues and are exposed to physiological environments, accurate characterisation of their elastic properties is essential for ensuring safe and reliable performance \cite{ref10,ref11}. However, accurately measuring the elastic properties of hydrogels requires isolating the pure elastic responses from the complex environment, e.g., under various hydrated and ionic conditions, and remains challenging.

In practical applications, hydrogels are typically exposed to complex aqueous environments, where the ionic strength of the surrounding solution influences their elastic response \cite{ref12}. Increasing ionic concentration reduces the osmotic pressure difference between the gel and its surroundings, thereby decreasing the equilibrium swelling of hydrogel \cite{ref13}. In addition, ion-specific interactions arising from different ionic species modify the hydrogen-bond network and the local water structure within the gel \cite{ref13,ref14}. As a result, the elastic behaviour is dependent on specific environmental conditions and that case-by-case consideration is often adopted \cite{ref15,ref16}. A constitutive model considering interstitial aqueous conditions is required.

Elasticity is influenced by both the surrounding ionic conditions and the degree of swelling. Flory–Rehner theory predicts that the stiffness of polymer networks decreases with increasing degree of swelling \cite{ref17,ref18}. Later studies have revealed a more complex stiffness evolution, characterised by an initial softening followed by entropic stiffening towards the end of swelling \cite{ref19,ref20}. Such behaviour can be described by a recent constitute model in which stiffness is expressed as a function of the volumetric deformation \cite{ref20}. However, this model account for the mechanical response of hydrogels primarily through swelling, without considering additional ionic effects.

To characterise hydrogel elasticity, several experimental setups can be adopted, among which the indentation testing is commonly used due to its experimental simplicity, high displacement resolution, and applicability to small and soft samples \cite{ref21,ref22,ref23}. In this method, the applied load-indentation depth relationship is interpreted using classical contact models \cite{ref24}. Most notably, Hertzian theory assumes that the measured force arises solely from elastic deformation under dry, non-adhesive contact conditions \cite{ref25}. However, fully swollen hydrogels are highly hydrated, and contact with the indenter surface leads to the spontaneous formation of a liquid bridge due to free water at the gel interface \cite{ref26,ref27}. The associated Laplace pressure introduces an additional capillary force during indentation \cite{ref28}, thereby violating the dry-contact assumption underlying Hertzian contact theory. The coupling between capillary forces and elastic deformation gives rise to elastocapillarity in hydrogels. Recent studies have examined interactions between soft hydrogel beads and reported the presence of capillary forces \cite{ref29}. However, in these studies, elastocapillary contributions have yet been explicitly separated from the measured elastic responses, and their influence on the quantified elastic properties remains unclear. 

To explore the elastocapillarity under the combined effects of swelling and environmental conditions, this work establishes an experimental setup for granular hydrogels to measure their elastic responses. The influence of capillary interactions on the measured elastic responses of hydrogels is examined and analysed for enriching a constitutive model with variable degrees of swelling and ionic concentration.

\section{Methodology}

\subsection{Materials}

\begin{figure}[ht]
\centering
  \includegraphics[width=8cm]{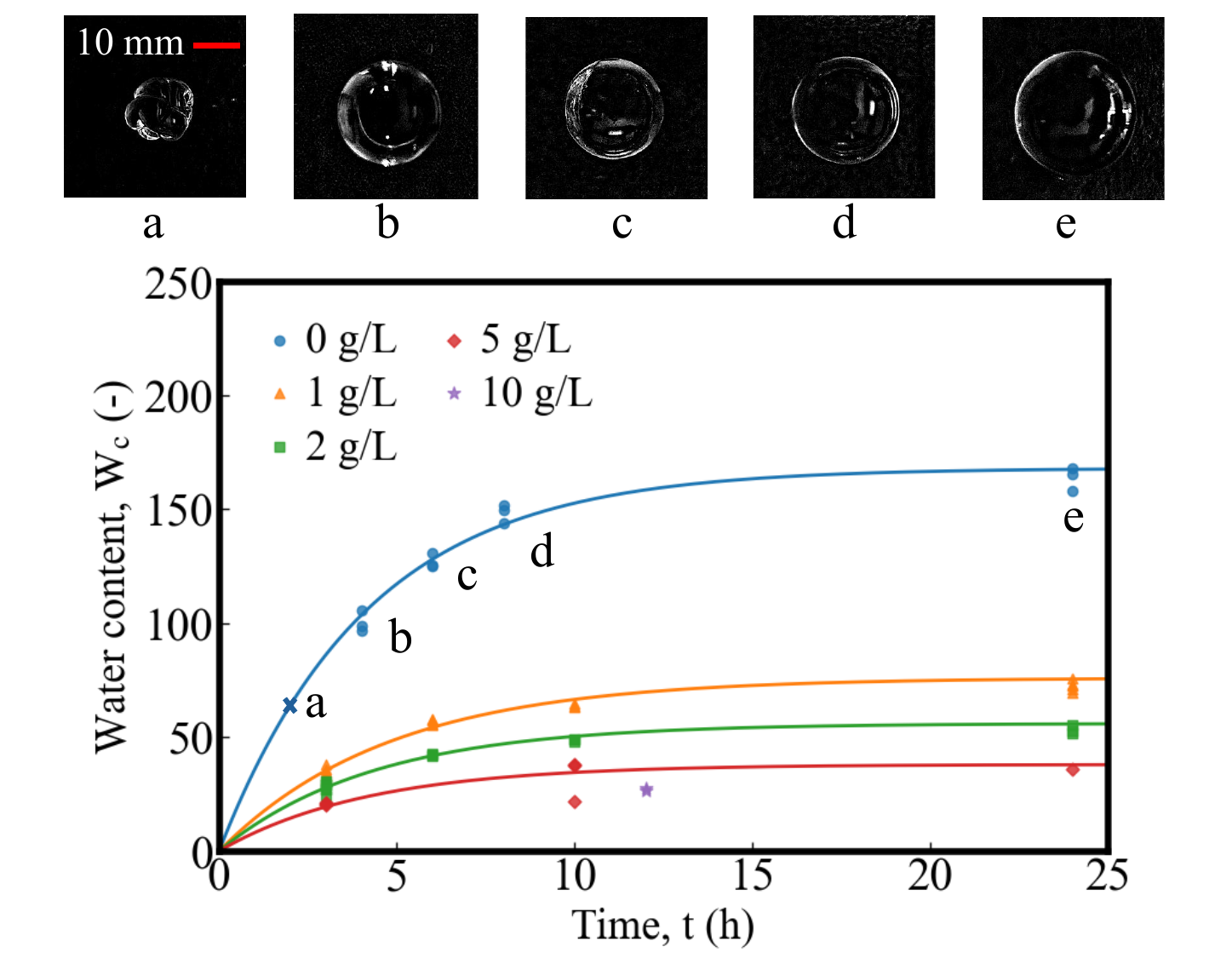}
  \caption{Temporal evolution of hydrogel water content under varying ionic conditions.}
  \label{Figure1}
\end{figure}

\begin{table}[ht]
\small
  \caption{Swelling radius and water content ranges of hydrogel spheres under different environments}
  \label{table1}
  \begin{tabular*}{\columnwidth}{@{\extracolsep{\fill}}lccccc}
    \hline
    \shortstack{Concentration of\\NaCl (g/L)} &
    \shortstack[c]{$R_{\min}$\\(mm)} &
    \shortstack[c]{$R_{\max}$\\(mm)} &
    \shortstack[c]{$W_{c,\min}$\\(--)} &
    \shortstack[c]{$W_{c,\max}$\\(--)} &
    \shortstack[c]{$k$\\($\mathrm{s}^{-1}$)} \\
    \hline
    0           & 7.4 & 9.0 & 100 & 168 & 0.244 \\
    $1 \pm 0.05$  & 5.2 & 7.0 & 30  & 76  & 0.229 \\
    $2 \pm 0.05$  & 4.8 & 6.2 & 25  & 56  & 0.254 \\
    $5 \pm 0.05$  & 4.3 & 5.5 & 20  & 38  & --    \\
    $10 \pm 0.05$ & 4.8 & 5.0 & 26  & 28  & --    \\
    \hline
  \end{tabular*}

\vspace{1mm}
{\footnotesize \textit{Notes:} $R_{\max}$ and $R_{\min}$ ($W_{c,\max}$ and $W_{c,\min}$) denote the radius (water content) of the hydrogel sphere corresponding to the fully swollen state obtained after 48~h of immersion and the moment when the hydrogel becomes spherical, meaning that no surface wrinkling or faceting is present. For sample groups of 5 and 10~g/L NaCl, the examinable water content ranges (from the moment being spherically shaped to fully swollen) are too narrow and will be excluded later for developing the constitutive model.}
\end{table}

Crosslinked polyacrylamide hydrogel spheres with dry diameters ranging from
$2.84 \pm 0.04$~mm were used as the model soft granular material. Samples were swollen
in sealed containers, with a single particle immersed in the desired solution for a given
duration (e.g., 3, 6, 10, and 24~h). Five ionic environments were considered: purified water,
and saline solutions with concentrations of 1, 2, 5, and 10~g/L NaCl. The swelling curves for
these samples are shown in Figure~\ref{Figure1}, with the corresponding trend curves following
$W_c(c,t)=W_{c,\max}(c)\left(1-\exp\!\left[-k(c)t\right]\right)$, where $W_{c,\max}(c)$ and $k(c)$ are the
maximum water content at fully swelling state and exponential coefficient, respectively, at a given ionic
concentration $c$. The water content is calculated by $W_c=(w-w_0)/w_0$, where $w$ and $w_0$ are the swelling
and dry weights of the hydrogel particle. The equilibrium swollen sizes and corresponding water content ranges
obtained under each condition are summarised in Table~\ref{table1}. Noted that, for sample groups of
5 and 10~g/L NaCl, the examinable water content ranges (from the moments being spherically shaped to fully swollen)
are too narrow and will be excluded later for developing the constitutive model.

\subsection{Experimental setup}

\begin{figure}[ht]
\centering
  \includegraphics[width=16cm]{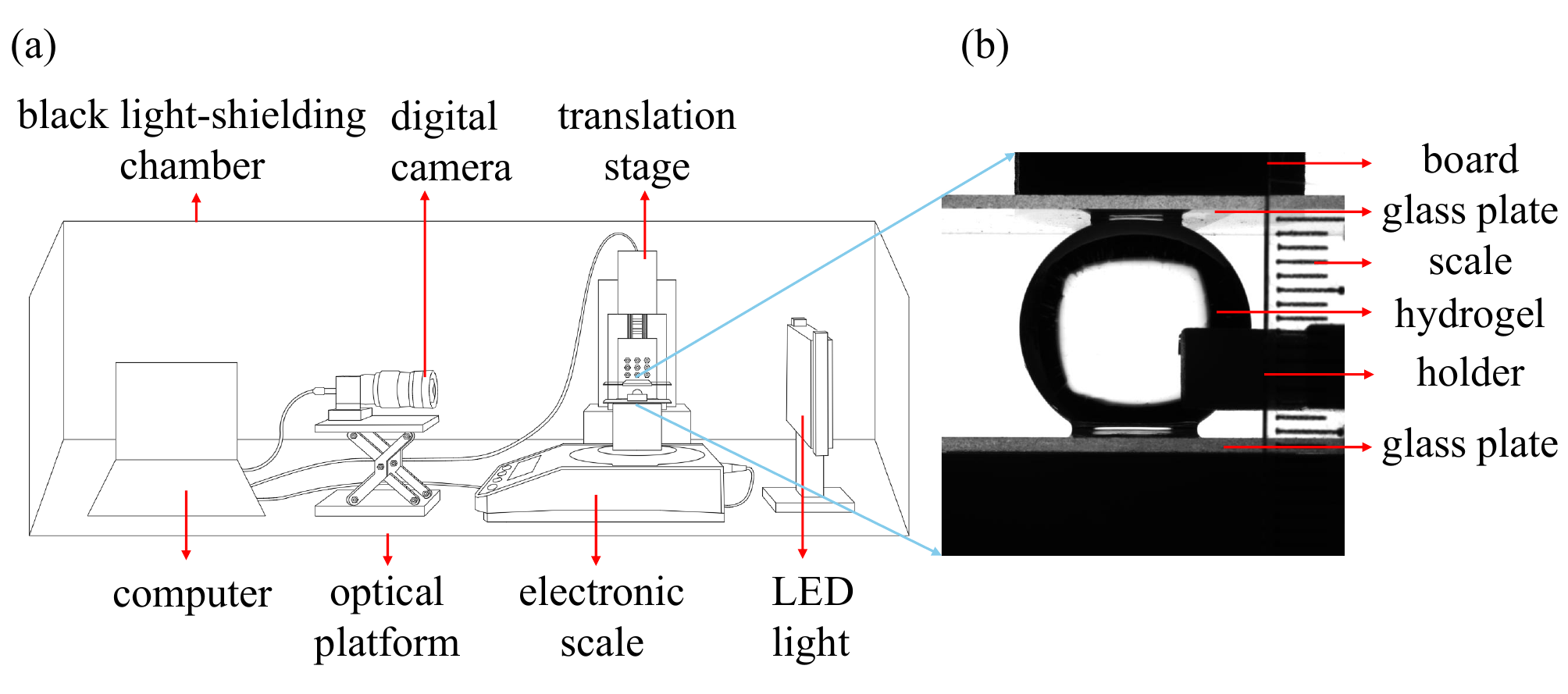}
  \caption{Experimental setup: (a) Overview of the custom-developed test setup (side view); (b) A typical experimental image captured by the camera.}
  \label{Figure2}
\end{figure}

A customised loading system was developed to monitor the response of hydrogel particles during compression, as shown in Figure~\ref{Figure2}. The system includes an electronic scale (A\&D, HR-250AZ, with an accuracy of 0.0001~g), a digital camera (DAHENG, MER2-2000-19U3M, with a spatial resolution of $2.4~\mu\mathrm{m}/\mathrm{pixel}$), a linear translation stage (Thorlabs, MTS25/M-Z8, with an increment accuracy of $\pm 0.8~\mu\mathrm{m}$), an optical platform (flatness: $\pm 0.05$~mm over $0.36~\mathrm{m}^2$), and a high-uniformity white LED light source (JSIONX, JS-DBL209-318). The entire loading system was enclosed within a black light-shielding chamber to eliminate the influence of ambient illumination and improve image consistency.

Two flat glass plates (sectional area: $25.4~\mathrm{mm}\times 76.2~\mathrm{mm}$; thickness: $\sim 1.1$~mm) were used in the experiments, one serving as the substrate and the other as the loading plate. The loading plate was mounted beneath a board attached to the linear translation stage. A 3D-printed holder was tailored to the size of the hydrogel to prevent lateral movement. A transparent ruler was affixed vertically to the side of the setup to serve as a scale reference in the captured images.

\subsection{Test procedures}

Before each test, the glass surfaces were rinsed with 96\% ethanol to ensure cleanliness and wettability consistency. The electronic scale was reset to zero and the LED light intensity was controlled to maintain consistent illumination across all tests. A single hydrogel particle was then placed onto the lower glass plate with the constraint from the 3D-printed sample holder. The camera was aligned with the centre of the liquid bridge and remained in focus throughout the loading process. The upper glass plate was subsequently lowered to a position just above the hydrogel without contact. The translation stage then initiated compression at a constant loading speed of 0.01~mm/s, with the maximum loading distance adjusted for each particle to approximately 5\% of its diameter. The electronic scale, translation stage and camera data were synchronised and recorded once per second via a Python script, ensuring a shared time reference and precise temporal alignment. It should be noted that this study was conducted within the practical swelling window corresponding to the minimum and maximum limits (as denoted in Table~\ref{table1}) that allow reliable mechanical measurements. Each set of experiments was conducted in triplicate to validate repeatability, and all experiments were conducted under an ambient temperature of $20 \pm 1~^{\circ}\mathrm{C}$ and humidity of 50\%--60\%.

\subsection{Image processing}

\begin{figure}[ht]
\centering
  \includegraphics[width=8cm]{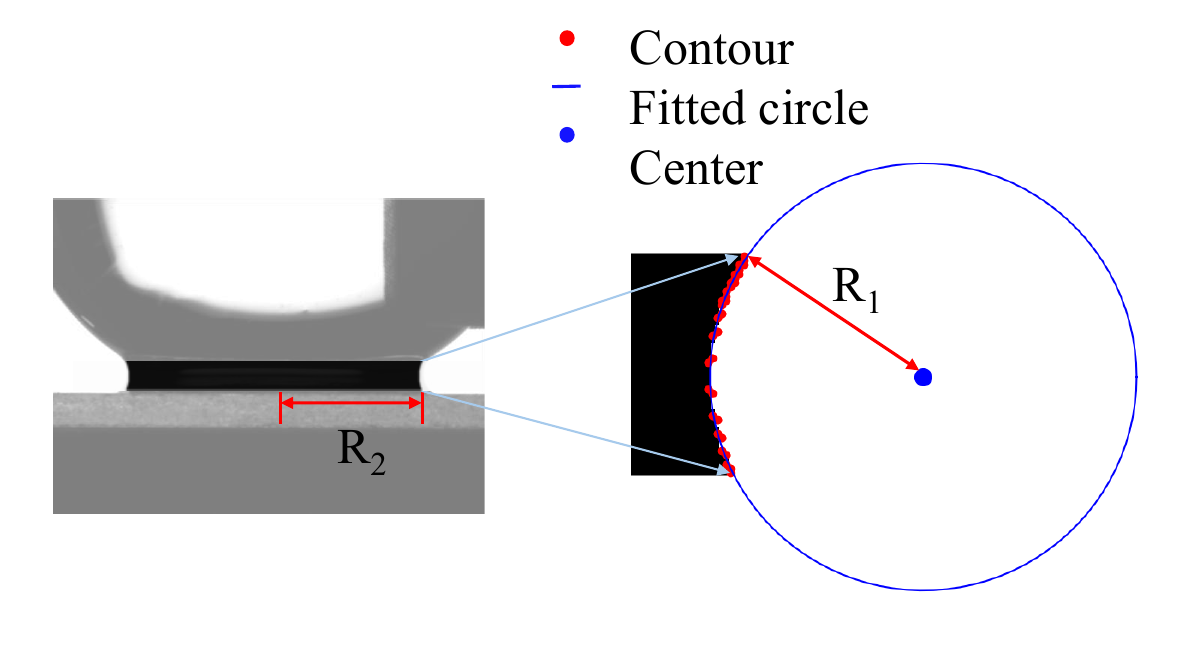}
  \caption{Image processing for extracting principal radii from the captured liquid bridge.}
  \label{Figure3}
\end{figure}

The captured images were pre-processed by stages of grayscale conversion, binarisation, and denoising using Gaussian blur. As shown in Fig.~\ref{Figure3}, the capillary bridge is isolated from the image, then boundary points on its interface were extracted and fitted by arcs to obtain the principal radii at each time step. The first principal radius ($R_1$) was calculated as the average fitted curvatures from the left and right sides of the liquid bridge. The second principal radius ($R_2$) was defined as the horizontal span between two fitted liquid profiles across the bridge neck. All radii were converted from pixel units to millimetres based on the scale reference in Figure~\ref{Figure2}.

\section{Results and Discussion}
\subsection{Effects of capillary interactions}

\begin{figure}[ht]
  \centering
  \includegraphics[width=16cm]{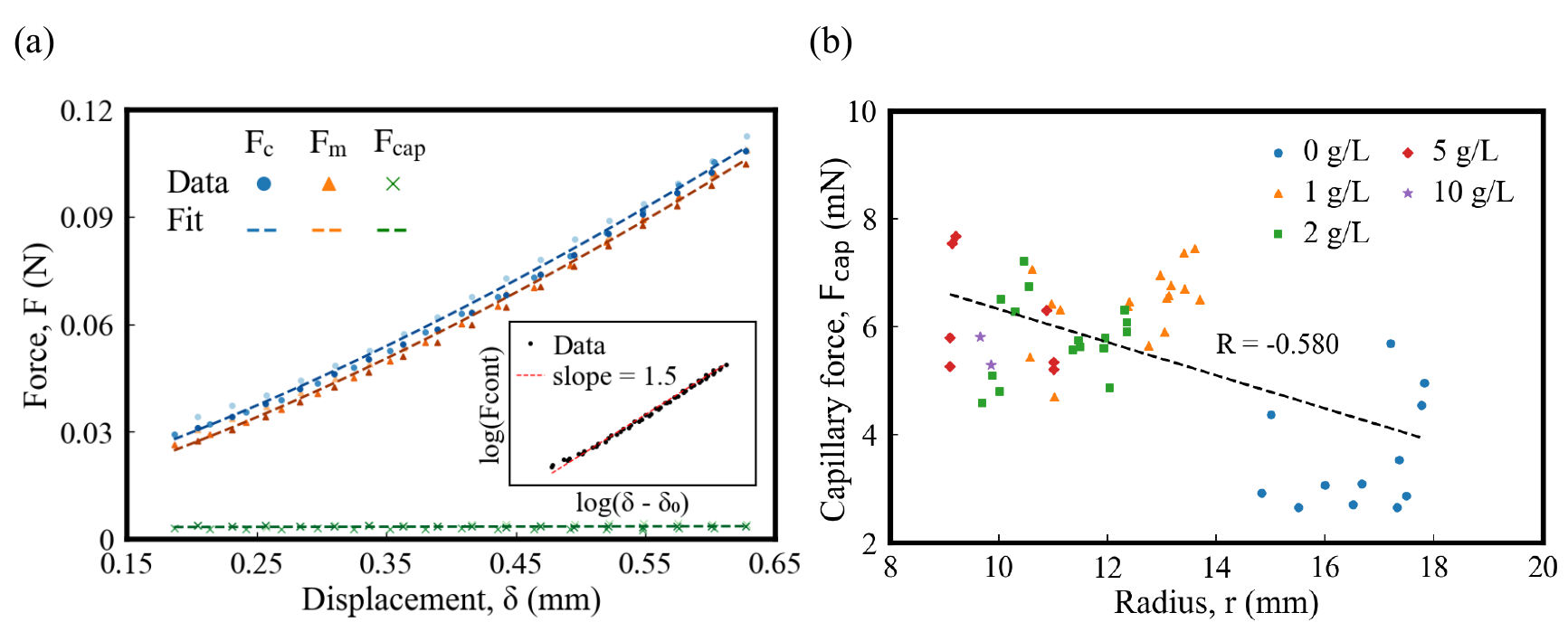}
  \caption{Effects of capillary interactions. (a) Typical force--displacement curves from replicate tests under purified water condition; (b) Measured capillary force versus particle radius.}
  \label{Figure4}
\end{figure}

Within this test configuration, the total contact force ($F_c$) acting on the hydrogel during compression comprises the sum of the measured force ($F_m$) and capillary force ($F_{\mathrm{cap}}$):
\begin{equation}
F_c = F_m + F_{\mathrm{cap}}.
\label{eq:force_decomposition}
\end{equation}
The measured force is obtained from the apparent mass recorded by the scale. The capillary force ($F_{\mathrm{cap}}$) is calculated by considering both the capillary pressure and interfacial tension as
\begin{equation}
F_{\mathrm{cap}} = \pi R_2^{\,2}\Delta p + 2\pi R_2 \gamma,
\label{eq:Fcap}
\end{equation}
where $\gamma$ denotes the surface tension, $\Delta p$ is the capillary pressure calculated as $\Delta p=\gamma(1/R_1+1/R_2)$, and $R_1$ and $R_2$ correspond to the principal radii of the liquid-bridge meniscus extracted from the experimental images.

The influence of salinity on interfacial parameters such as contact angle and surface tension is minimal within the tested concentration range. Previous studies have shown that NaCl concentrations below 10~g/L change these interfacial quantities by less than 0.05~mN/m \cite{ref30,ref31}. In the following analysis, we treat the surface tension as a constant ($\gamma = 72.8$~mN/m) and measure an equilibrium contact angle of $60 \pm 5^{\circ}$ under the controlled laboratory conditions.

As typical examples, the force--displacement curves obtained from three representative hydrogel samples tested under identical conditions (0~g/L saline, fully saturated) are shown in Figure~\ref{Figure4}a. The measured forces ($F_m$), contact forces ($F_c$) and capillary forces ($F_{\mathrm{cap}}$) each show highly consistent and nearly overlapping trends across repeated tests, confirming that the experimental platform yields stable and repeatable measurements under controlled conditions. Note that the initial points on these curves are non-zero because the first force measurement already contains the self-weight of the hydrogel, and the displacement has been adjusted by removing a threshold value (discussed later in this section). The capillary force generated by the liquid bridge remains almost constant during the tests and has noticeable effects highlighted by the difference between the measured forces ($F_m$) and contact forces ($F_c$).

The particle size directly influences the principal curvatures of the liquid bridge and therefore affects the magnitude of the capillary force. Figure~\ref{Figure4}b presents the dependence of the mean capillary force on particle radius. Here, mean values of $F_{\mathrm{cap}}$ are reported given that temporal variations remain small, as shown in Figure~\ref{Figure4}a. Across all saline conditions, the capillary force lies within 2--8~mN, indicating that the liquid-bridge force remains of similar order of magnitude for all samples. The trend line reveals a slight decrease in $F_{\mathrm{cap}}$ with increasing particle size, implying that smaller hydrogels sustain larger capillary forces during the tests.

\subsection{Effective Young's modulus}

\begin{figure}[ht]
\centering
  \includegraphics[width=16cm]{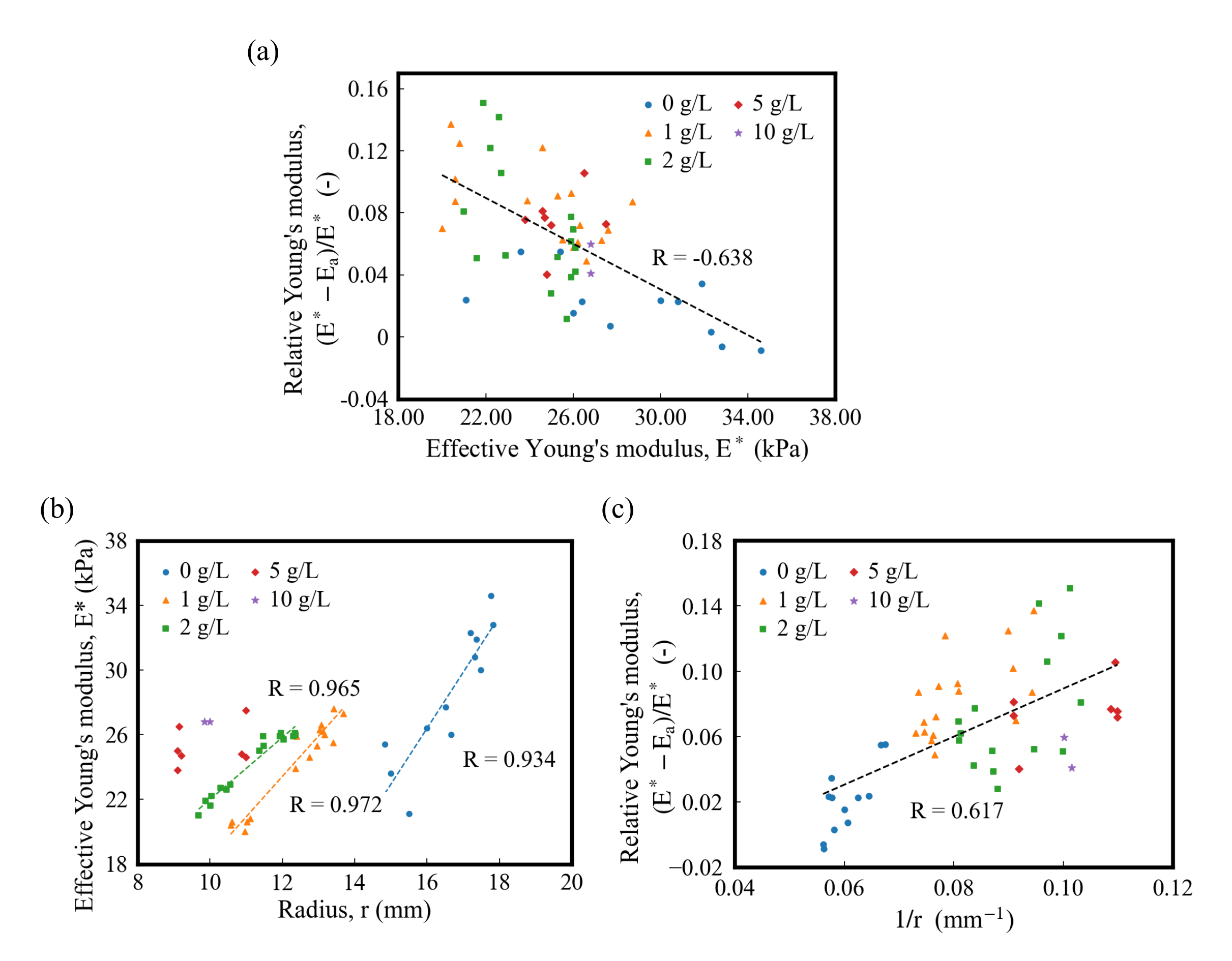}
  \caption{Size-dependent capillary effects. (a) Relative deviation of Young’s modulus versus the effective Young’s modulus; (b) Effective Young’s modulus as a function of particle radius; (c) Size-dependence of relative Young’s modulus.}
  \label{Figure5}
\end{figure}

As demonstrated above, capillary forces introduce a size-dependent discrepancy between the measured force ($F_m$) and the actual contact force ($F_c$) in wet particle compression. To quantify the resulting discrepancy, we separately evaluate the apparent Young's modulus ($E_a$) and the effective Young's modulus ($E^{*}$) from the measured force ($F_m$) and contact force ($F_c$), respectively, based on Hertzian contact theory:
\begin{equation}
F=\frac{4}{3}E\,R^{1/2}\left(\delta-\delta_0\right)^{3/2},
\label{eq:hertz_shifted}
\end{equation}
where $E$ is the Young's modulus, $R$ is the hydrogel radius, and $\delta$ is half of the loading-plate displacement relative to the initial position. Here $\delta_0$ is the threshold displacement that marks the onset of mechanical contact at the bottom plate; it is obtained through curve fitting and is independently determined for $F_c$ and $F_m$.

Figure~\ref{Figure5}a shows that the relative Young's modulus, $(E^{*}-E_a)/E^{*}$, increases as $E^{*}$ decreases, indicating that samples with lower stiffness are more affected by the bias induced by capillary forces.

The size dependence of the measured effective Young's modulus ($E^{*}$) under different saline conditions is shown in Figure~\ref{Figure5}b. For the 0, 1, and 2~g/L samples, $E^{*}$ increases with particle radius due to different water content, with each salinity group exhibiting a clear positive correlation within the respective water-content range $[W_{c,\min},\,W_{c,\max}]$. Moreover, the slope of the fitted trend line increases with salinity, suggesting that hydrogels swollen in higher-salt environments stiffen more rapidly with increasing water content. At higher salinities (5 and 10~g/L), the data span a much narrower water-content range and therefore do not show a pronounced trend.

To isolate the role of capillary effects in the observed size dependence, the relative Young's modulus $(E^{*}-E_a)/E^{*}$ is plotted as a function of $1/r$ in Figure~\ref{Figure5}c, since the key difference between $F_m$ and $F_c$ scales with $\gamma/r$, consistent with elastocapillary scaling. The positive correlation reveals that the apparent modulus increasingly underestimates the effective modulus as particle size decreases, with deviations reaching up to approximately 15\%. This indicates that for smaller particles, neglecting capillary forces causes a larger error in the inferred Young's modulus.

\subsection{Constitutive model for effective modulus}

\begin{figure}[ht]
\centering
  \includegraphics[width=8cm]{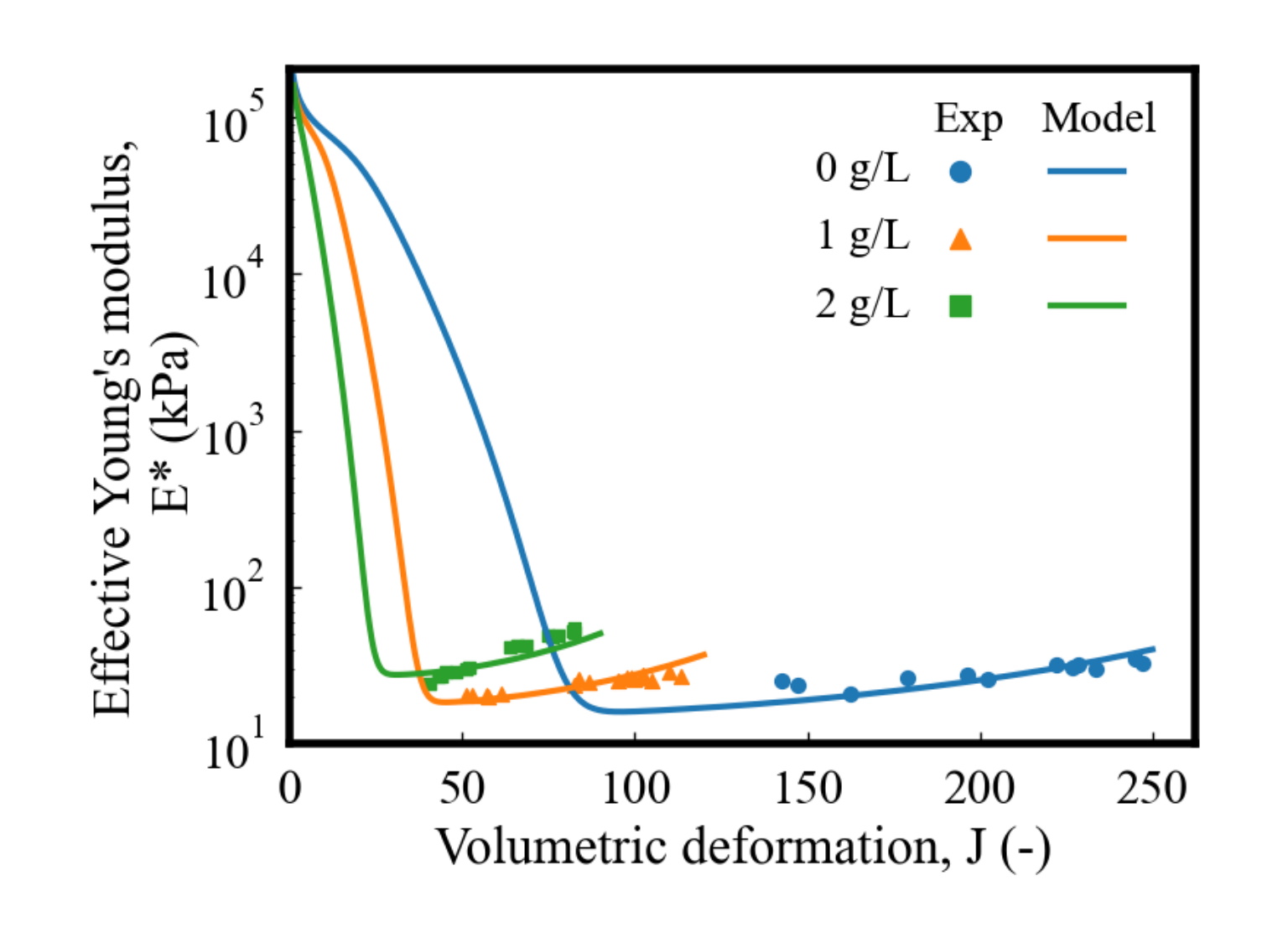}
  \caption{Comparison between the measured modulus and model predictions as a function of volumetric deformation.}
  \label{Figure6}
\end{figure}

The evolution of $E^{*}$ with swelling can be further interpreted using the theoretical framework of Brighenti \textit{et al.}\cite{ref20} which describes hydrogel elasticity through a molecular--statistical formulation. In this model, the effective Young's modulus is expressed as
\begin{align}
E &= J^{-1/3}\Bigg[
\frac{NkT\left(5\rho_0^{4}-16\rho_0^{2}+15\right)}{5\left(1-\rho_0^{2}\right)^{2}}
+\frac{N_g\alpha}{2}\left(\frac{n_g}{n}-1\right)k_{\mathrm{hb}}\,l_{\mathrm{hb}}^{2}
\Bigg],
\label{eq:brighenti_E}
\end{align}
where $J$ is the volumetric deformation defined as $V/V_0=(R/R_0)^3$. $N$ and $N_g$ denote the chain densities in the swollen and fully swollen limits, respectively, with $N_g$ defined as $G/kT$, where $G$ is the shear modulus of the fully swollen network and $kT$ is the thermal energy. The parameter $\alpha$ represents the number of intermolecular hydrogen bonds per repeat unit. The term $\rho_0$ represents the ratio between the end-to-end distance and the contour length of a polymer chain in the swollen state. The evolution of repeat units per chain is described by $n(J)$, which follows the bond-dissociation kinetics:
\begin{align}
n(J) &= \frac{n_g n_d}{2}\,\operatorname{erf}\!\left[\gamma_T\left(J-J_c\right)\right]
+ \frac{n_g n_d}{2},
\label{eq:nJ_erf}
\end{align}
where $n_d$ is the dry-state chain density, $J_c$ is a critical deformation that defines the transition point, and $\gamma_T$ is the rate parameter controlling the transition from softening to stiffening. Material constants such as the effective H-bond length $l_{\mathrm{hb}}$ and single-bond stiffness $k_{\mathrm{hb}}$ introduce the energetic contribution arising from hydrogen-bond rupture.

This theoretical framework has been developed and validated using experimental data for hydrogels swelling in water.\cite{ref20} Under variable aqueous conditions, the material parameters sensitive to the ionic concentration can be identified as $G$, $\gamma_T$, and $J_c$. These parameters describe the material properties at fully swollen conditions and the swelling kinematics under variable osmotic conditions. A comparison between the model predictions and the experimental measurements is presented in Figure~\ref{Figure6}. The parameters adopted in this study are $kT=4.14\times 10^{-21}$~J, $l_{\mathrm{hb}}=0.3$~nm, $k_{\mathrm{hb}}=10$~nN/nm, $n_d=5$, $n_g=35$, and $\alpha=18$. Salinity-dependent behaviour is introduced through the fitted parameters in Table~\ref{table2}. Increasing NaCl concentration leads to higher $G$ and $\gamma_T$, along with reduced $J_c$, consistent with a stiffer and less extensible polymer network under stronger ionic screening.

\begin{table}[ht]
\small
\caption{Salinity-dependent fitted parameters used in the constitutive model}
\label{table2}
\begin{tabular*}{\columnwidth}{@{\extracolsep{\fill}}cccc}
\hline
\shortstack[c]{Concentration of\\NaCl (g/L)} &
\shortstack[c]{$G$\\(kPa)} &
\shortstack[c]{$J_c$\\(--)} &
\shortstack[c]{$\gamma_T$\\(--)} \\
\hline
0 & 15.0 & 25 & 0.08 \\
1 & 15.5 & 24 & 0.10 \\
2 & 23.0 & 21 & 0.14 \\
\hline
\end{tabular*}

\vspace{1mm}
{\footnotesize \textit{Notes:} $G$, $J_c$, and $\gamma_T$ are salinity-dependent parameters obtained by fitting the constitutive model to the experimental data. All other model parameters are kept constant across salinity conditions as listed in the main text.}
\end{table}

The constitutive model successfully reproduces the key features of swelling-dependent stiffness, including the pronounced modulus reduction at early swelling followed by the gradual increase at larger volumetric deformation. In the work of Brighenti \textit{et al.},\cite{ref20} the stiffening branch of the model was supported by a dataset obtained under a single aqueous condition. In the present study, the 0--2~g/L datasets span a broader swelling range and include multiple conditions that enter the stiffening regime, allowing a more systematic comparison with the model. Through quantitative comparison, good agreement between the experimental datasets and model predictions is observed in Fig.~\ref{Figure6}, with relative errors below 15\%.

Across the three salinity levels (0--2~g/L), as salinity increases, the softening regime becomes steeper and the modulus reaches its minimum at a smaller volumetric deformation, indicating that higher ionic strength accelerates the initial network dilution. In the stiffening regime, the slope of the ascending branch increases with salinity, showing that networks swollen under higher-salt conditions recover stiffness more rapidly once chain stretching becomes dominant. Furthermore, for any given $J$ within the overlapping deformation range, hydrogels exposed to higher salinity consistently exhibit larger $E^{*}$, consistent with increased polymer chain density and reduced extensibility imposed by ionic screening.

\subsection{Discussion}

The consistency observed in the low-salinity datasets shown in Figure~\ref{Figure6} provides a basis for applying the modified model to interpret or extrapolate behaviour at higher salinity. Note that at higher salinity conditions, due to the presence of higher osmotic pressure, the ionic concentration may not reach equilibrium during swelling.\cite{ref30} The 5 and 10~g/L datasets contain fewer measurable swelling states and therefore do not permit direct fitting. However, their magnitudes and qualitative trends remain compatible with the model evolution. These observations suggest that the constitutive model may be employed to predict modulus evolution in salinity regimes where experimental access is limited, offering a practical route for estimating stiffness under conditions that fall outside the experimentally accessible swelling window.

The current experimental setup may be limited to testing convex-shaped samples, such as these granular hydrogels, allowing direct optical observation of capillary forces. For other indentation tests, the liquid bridges formed are typically obscured during the test, and extra care should be taken when extracting elastic properties from such measurements.

Although the present study has focused on the elastic response of swollen hydrogels, it is noted that hydrogels are intrinsically viscoelastic materials whose behaviour involves time-dependent deformation, solvent migration, and energy dissipation. Viscous relaxation and potential fracture may influence the contact response under different loading rates or larger deformations. These effects were beyond the scope of the current study but represent important directions for future research.

\section{Conclusions}
This study investigated the coupled elastic and capillary response of swollen hydrogel particles during compression tests against a rigid surface. The custom-built experimental setup enabled direct and stable observation of liquid-bridge evolution, allowing capillary forces to be quantified and separated from the mechanical load. The results demonstrate that neglecting capillary forces leads to a systematic underestimation of the effective elastic modulus, particularly for smaller or softer particles. A size-dependent deviation between the apparent and capillary-corrected modulus was identified, and the measured variation was found to be consistent with elastocapillary scaling. Furthermore, the dependence of stiffness on water content and salinity was characterised, and the experimental results showed good correspondence with a modified theoretical model, supporting its applicability across different swelling regimes and ionic conditions.

\section*{Author contributions}
Jiayin Zhao: Methodology, Software, Investigation, Validation, Formal analysis, Data curation, Visualization, Writing – original draft. Haiyi Zhong: Methodology, Formal analysis, Visualization, Writing – review \& editing, Supervision, Resources. Yixiang Gan: Conceptualization, Methodology, Supervision, Resources, Formal analysis, Writing – review \& editing, Visualization.

\section*{Conflicts of interest}
There are no conflicts to declare.

\section*{Data availability}
The data that support the findings of this study are available from the corresponding author upon reasonable request.

\section*{Acknowledgements}
The authors acknowledge the support from the School of Civil Engineering, The University of Sydney. We also thank colleagues for helpful discussions and technical assistance.

\bibliographystyle{unsrtnat} 
\bibliography{ref}
\end{document}